\documentclass[runningheads]{llncs}

\usepackage[T1]{fontenc}
\usepackage{indentfirst}
\usepackage{listings}
\usepackage{algorithm}  
\usepackage{algorithmicx}
\usepackage{algpseudocode}
\usepackage{amsmath}
\usepackage{siunitx}
\usepackage{amssymb}
\usepackage{colortbl}
\usepackage{multirow}
\usepackage{subfigure}
\usepackage{graphicx}
\begin{document}
\title{A Linear Combination-based Method to Construct Proxy Benchmarks for Big Data Workloads}
\titlerunning{Linear Combination-based Proxy Benchmark Construction Method}
\author{Yikang Yang\inst{1,2} \and
Lei Wang\inst{1,2} \and
Jianfeng Zhan\inst{1,2}}
\authorrunning{Y. Yang, L. Wang et al.}
\institute{Institute of Computing Technology, Chinese Academic of Science \\
\email{\{yangyikang23s,wanglei\_2011,zhanjianfeng\}@ict.ac.cn}
\and
University of Chinese Academy of Sciences, China}
\maketitle
\begin{abstract}
\par
During the early stages of CPU design, benchmarks can only run on simulators to evaluate CPU performance. However, most big data component benchmarks are too huge at the code size scale, which causes them to be unable to finish running on simulators at an acceptable time cost, as running on the simulators requires more than 100X-1000X time consumed than running on the physical platform.
Moreover, big data benchmarks usually need complex software stacks to support their running, which is hard to be ported on the simulators. Proxy benchmarks, without long running times and complex software stacks, have the same micro-architectural metrics as real benchmarks, which means they can represent real benchmarks' micro-architectural characteristics. Therefore, proxy benchmarks can replace real benchmarks to run on simulators to evaluate the CPU performance. 
\par
The biggest challenge is how to guarantee that the proxy benchmarks have exactly the same micro-architectural metrics as real benchmarks when the number of micro-architectural metrics is very large. To deal with this challenge, we propose a linear combination-based proxy benchmark generation methodology that transforms this problem into solving a linear equation system. We also design the corresponding algorithms to ensure the linear equation is astringency, which means that although sometimes the linear equation system doesn't have a unique solution, the algorithm can find the best solution by the non-negative least square method.
\par
We generate fifteen proxy benchmarks and evaluate their running time and accuracy in comparison to the corresponding real benchmarks for Mysql and RockDB. On the typical Intel Xeon platform, the average running time is 1.62 seconds, and the average accuracy of every micro-architectural metric is over 92\%, while the longest running time of real benchmarks is nearly 4 hours. We also conduct two case studies that demonstrate that our proxy benchmarks are consistent with real benchmarks both before and after prefetch or Hyper-Threading is turned on. 
\end{abstract}
\section{Introduction}
\par
In recent years, big data systems, including traditional relational databases, non-relational databases, and distributed data management systems, have been making an increasingly significant contribution to the development of economy~\cite{DatabaseResearch,MongodbvsOracle,LargeScaleDataAnalysis,SQLvsNoSQL}. The CPU requires more advanced designs to enhance performance, while benchmarks are important tools for evaluating CPU performance. Compared to traditional benchmarks like SPECCPU~\cite{SPECCPU2006} and PARSEC~\cite{PARSEC}, big data benchmarks like CloudSuite~\cite{CloudSuite} and BigDataBench~\cite{gao2018bigdatabench,BigDataBenchWebSearchEngines,BigDataBenchInternetServices} can provide
a more accurate evaluation of the CPU performance in processing big data tasks. 
\par
During the early stages of CPU design, the validity and effectiveness of many designs have to be verified on simulators due to the heavy cost of designing and implementing a CPU system. However, big data benchmarks can not run on simulators because of prohibitively heavy time costs and the lack of supporting  complex software stacks on the CPU simulators.

\par
Proxy benchmarks are workloads used to replace real benchmarks for evaluating CPU performance. Compared to real big data benchmarks, proxy benchmarks have a short running time and don't need to port complex software stacks on simulators. Moreover, they have the same micro-architectural metrics as real big data benchmarks, which means they can represent real benchmarks' micro-architectural characteristics. Han et al. propose Cloudmix~\cite{Cloudmix} to construct proxy benchmarks for cloud systems, but these proxy benchmarks don't have similar metrics in cache behavior, branch prediction, and instruction mix. Panda et al. propose PerfProx~\cite{PerfProx} methodology to construct proxy benchmarks for database applications, but these proxy benchmarks don't align real benchmarks' micro-architectural metrics directly, which means there are gaps with real benchmarks in terms of micro-architectural metrics. Gao et al. propose a data motif-based proxy benchmark generation methodology~\cite{data-motif,data-motif-proxy}, but this methodology requires the source codes of real benchmarks, which are sometimes not available. 
\par
The biggest challenge for constructing proxy benchmarks is how to guarantee the proxy benchmarks have exactly the same micro-architectural metrics as real benchmarks when the number of micro-architectural metrics is very large. Because there are correlations between the metrics, they can mutually influence each other, making it difficult to align them simultaneously. Previous work primarily focused on aligning small-scale metrics, and there is no previous work addressing the alignment of large-scale metrics. We propose a linear combination-based proxy benchmark generation methodology to deal with the challenge.
\par
Our contributions are three-fold as follows:
\begin{itemize}
    \item[$\bullet$] We propose a linear combination-based proxy benchmarks generation methodology that transforms the problem of constructing a proxy benchmark into solving a linear equation system by a non-negative least square method. This methodology can easily be expanded by just adding some equations into the system to deal with large-scale metrics.
    \item[$\bullet$] We generate fifteen proxy benchmarks for real big data benchmarks. The evaluation results demonstrate that the average accuracy of each micro-architectural metric is over 92\% while the average running time is 1.62 seconds.
    \item[$\bullet$] We use the fifteen proxy benchmarks to conduct two case studies, which demonstrate that our proxy benchmarks are consistent with real benchmarks both before and after prefetch or Hyper-Threading is turned on.
\end{itemize}
\par
The rest of this paper is organized as follows. Section~\ref{section: Proxy Benchmark Generation} provides the proxy benchmark generation methodology. Section~\ref{section: Evaluation} conducts evaluations for proxy benchmarks. Section~\ref{section: Case Studies} presents two case studies for micro-architecture configuration settings. Section~\ref{section: Related Work} discusses the related work. Section~\ref{section: Conclusion} draws conclusions.
\section{Proxy Benchmark Generation Methodology}
\label{section: Proxy Benchmark Generation}
\subsection{Problem Description}
Proxy benchmarks constructed must have the same micro-architectural metrics as real benchmarks, while the metrics are listed in Table \ref{Micro-architectural Metrics}. The metrics can be divided into five categories, including processor performance, branch prediction, cache behavior, TLB behavior, and instruction mix. 
Perf~\cite{IntelDeveloperManul,Perf}, a hardware event counter, is used by us to collect these metrics. 
Because there are strong correlations between these metrics and their number is high, they can mutually influence each other, making it difficult to align them simultaneously.
Except for the requirement of micro-architectural metrics, proxy benchmarks must have short running time and don't need complex software stacks.
\begin{table}
\centering
\caption{Micro-architectural Metrics}
\begin{tabular}{|l|l|l|}
\hline
Category & Metric Name & Description \\
\hline
Processor Performance & CPI & cycles per instruction \\
\hline
Branch Prediction & Branch Miss & Branch misprediction rate \\
\hline
\multirow{4}*{Cache Behavior} & L1 DCache Miss & L1 data cache miss rate \\
\cline{2-3}
~ & L1 ICache Miss & L1 instruction cache miss rate \\
\cline{2-3}
~ & L2 Cache Miss & L2 cache miss rate \\
\cline{2-3}
~ & L3 Cache Miss & L3 cache miss rate \\
\hline
\multirow{2}*{TLB Behavior} & DTLB Miss & data TLB miss rate \\
\cline{2-3}
~ & ITLB Miss & instruction TLB miss rate \\
\hline
Instruction Mix & Instruction ratios & \begin{tabular}[c]{@{}l@{}}
Ratios of load, store, branch, floating point, \\integer and vector instructions
\end{tabular} \\
\hline
\end{tabular}
\label{Micro-architectural Metrics}
\end{table}
\par
\begin{figure}[htb]
    \centering 
    \includegraphics[scale=0.22]{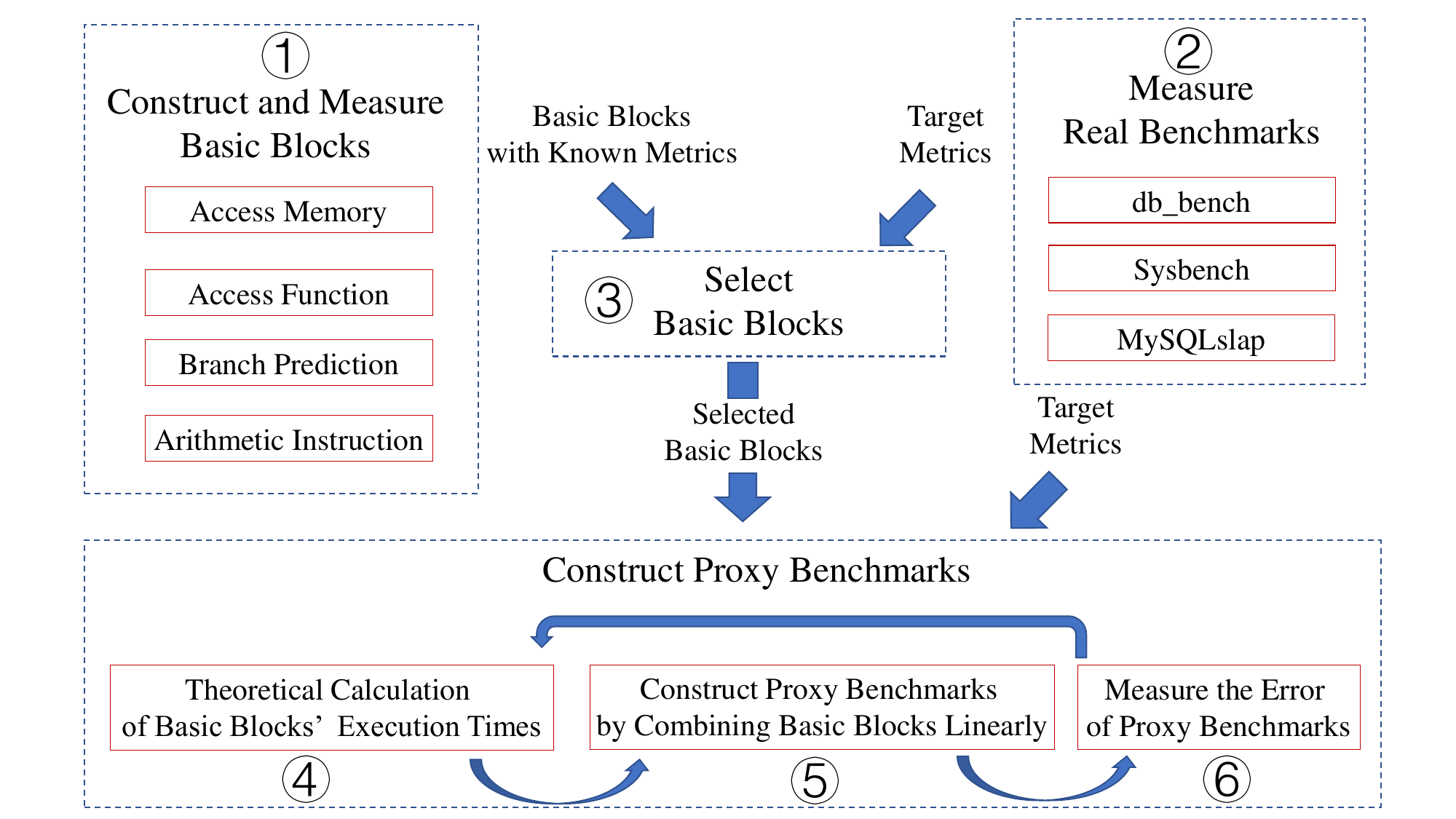}
    \caption{Proxy Benchmark Generation Methodology.}
    \label{fig: Proxy Benchmark Generation Methodology}
\end{figure}

We propose a linear combination-based proxy benchmark generation methodology illustrated in Fig. \ref{fig: Proxy Benchmark Generation Methodology}. Firstly, we measure the micro-architectural metrics of each real big data benchmark. These metrics will serve as the target metrics for our proxy benchmarks. Next, we select appropriate program fragments \textemdash basic blocks from the basic block set based on the target metrics. Finally, we combine these selected basic blocks linearly to construct proxy benchmarks. To reduce the gap between the real benchmarks and the proxy benchmarks, we adjust the execution times of each basic block in multiple rounds of iteration. This process ensures that the proxy benchmark closely resembles the real benchmark in terms of micro-architectural metrics. 
\par
\subsection{Basic Block}
\par
\begin{figure}
\centering
\includegraphics[scale=0.6]{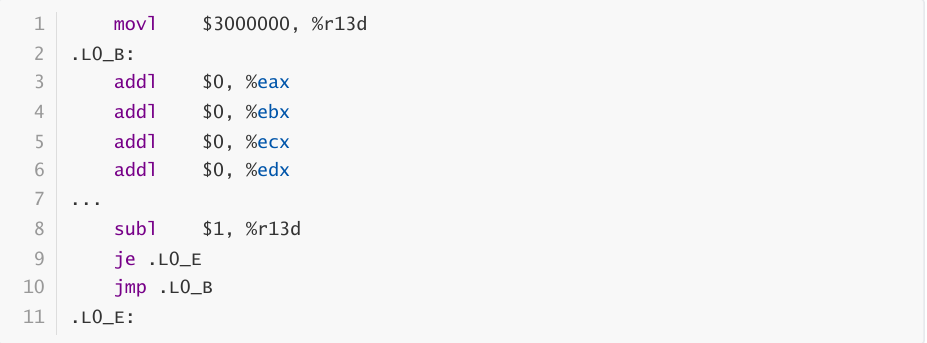}
\caption{Example of Basic Block.} \label{fig: Example of Basic Block}
\end{figure}

We define a basic block as a program fragment consisting of a specific series of assembly instructions, while their structures can be divided into interior and exterior. The exterior is a loop, which can control the execution times of the basic block by changing the number of iterations. The interior structure is composed of a series of specific instructions, which determines the micro-architectural metrics of the basic block. Fig. \ref{fig: Example of Basic Block} presents an example of a basic block, in which the interior comprises several add instructions, and the exterior structure controls it to execute 300,000 times. Based on the differences in the interior instruction series, we have constructed a total of four distinct types of basic blocks.
\par
\textbf{Memory Access Basic Block}. This kind of basic block primarily focuses on accessing memory at a fixed distance. The fixed distance between two adjacent memory accesses is denoted as STRIDE. An increase in STRIDE can result in a decrease in spatial locality, leading to higher miss rates in L1 DCache and DTLB. Furthermore, the increased miss rates in L1 DCache can potentially impact the miss rates of L2 Cache and L3 Cache. 
Therefore, by adjusting the STRIDE value, we can construct a set of access memory basic blocks with varying miss rates in L1 DCache, L2 Cache, L3 Cache, and DTLB. 
\par
\textbf{Access Function Basic Block}. This kind of basic block focuses on accessing functions at a fixed distance. These functions have different addresses and are arranged sequentially in memory. The fixed distance between two adjacent function accesses is denoted as STRIDE. A larger STRIDE value weakens the temporal locality, resulting in increased miss rates in L1 ICache and ITLB. The increased miss rates in L1 ICache can potentially impact the miss rates of L2 Cache and L3 Cache. 
Therefore, by adjusting the STRIDE value, we can construct a set of access function basic blocks with varying miss rates in L1 ICache, L2 Cache, L3 Cache, and ITLB. 
\par
\textbf{Branch Prediction Basic Block}. This kind of basic block utilizes a random number and a threshold value for branch prediction. It will generate a random number R between 0 and 1024 and compares R with a pre-set THRESHOLD. If R is bigger than THRESHOLD, a branch jump will happen; otherwise, it will not. As THRESHOLD approaches 512 (half of 1024), the branch jump becomes more random, making branch prediction more difficult and increasing the branch misprediction rate. 
By adjusting THRESHOLD, we can create a set of branch prediction basic blocks with a wide distribution of branch misprediction rates.
\par
\textbf{Arithmetic Instruction Basic Block}. The interiors of these basic blocks primarily consist of various arithmetic instructions. These basic blocks are designed to showcase the differences in execution speeds among different arithmetic instructions. Specifically, add and sub instructions are typically faster, mul instructions are slower, and div instructions are the slowest. By designing a set of basic arithmetic instruction blocks with different combinations and sequences of these instructions, we can create basic blocks with varying Cycles Per Instruction (CPI) values. 
\subsection{Linear Combination Method}
\label{Linear Combination Method}
\par
We generate proxy benchmarks by linearly combining basic blocks. The linear combination involves both connection and scaling, with their definitions stated as Definition \ref{def: Connection} and Definition \ref{def: Scaling}. 
\begin{definition}
    \label{def: Connection}
    \textbf{Connection} refers to the act of linking two program fragments in a specific order. If we denote the higher-order program fragment as  $P_1$ and another $P_2$, we will represent the connected program fragment as $P_1 + P_2$.
\end{definition}
\begin{definition}
    \label{def: Scaling}
    \textbf{Scaling} refers to altering the execution times of a basic block. If a basic block $P$ is scaled by $k$ times to create a modified basic block $P^{'}$, we will use $k \cdot P$ to represent $P^{'}$
\end{definition}
\par
During the execution of every program, various hardware events will occur on the microprocessor, including the misses and accesses of Cache at all levels, the misses and accesses of DTLB and ITLB, the branch predictions and the branch mispredictions, etc. If 
we use $f(P)$ to represent the number of times a hardware event $f$ occurs during the execution of program $P$, then Equation \eqref{add} and Equation \eqref{mul} will hold true under ideal conditions.
\begin{equation}
f(P_1+P_2)=f(P_1)+f(P_2) \label{add}
\end{equation}
\begin{equation}
f(k \cdot P)=k \cdot f(P) \label{mul}
\end{equation}
\par
\begin{figure}
\centering
\includegraphics[scale=0.25]{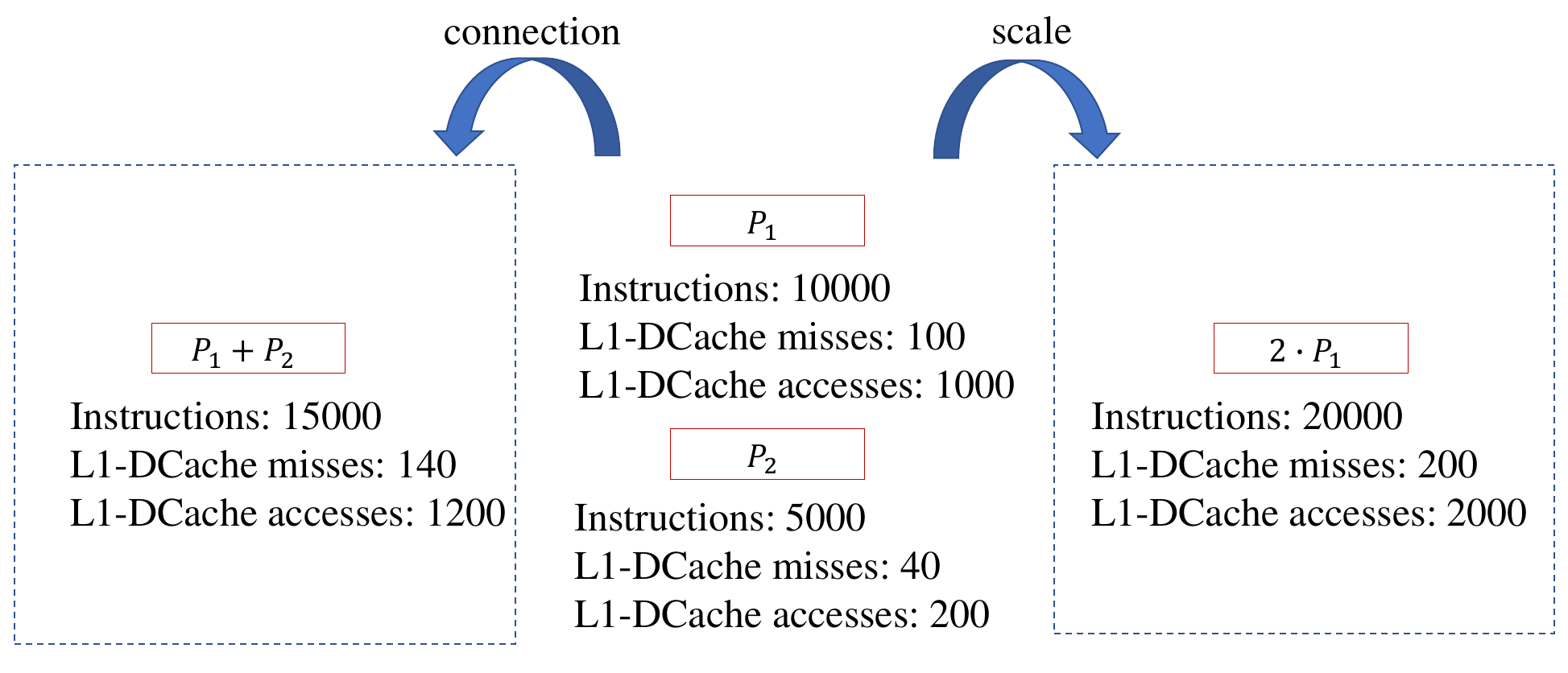}
\caption{Explanation of Linear Combination.} \label{Explanation of Linear Combination}
\end{figure}

Therefore, as Fig. \ref{Explanation of Linear Combination} describes, when two basic blocks are connected together, the occurrence times of each hardware event will equal the original sum, and when a basic block is scaled by 2 times, the occurrence times of each hardware event will double.
\par
We measure the occurrence times of all hardware events by running all basic blocks $N_0$($N_0=10^7$) times. For a proxy benchmark, we construct it with basic blocks $P_1, P_2,..,P_m$ and denote the execution times of the kth basic block as $N_k$. During the execution of the proxy benchmark, for a hardware event $f$, it will occur $\sum_{j=1}^{m}f(P_j) \cdot \cfrac{N_j}{N_0} $ times. Additionally, every micro-architectural metric can be calculated using the quotient of two hardware events. For example, the miss rate of L1 DCache can be calculated by dividing the L1 DCache misses by the L1 DCache accesses. Similarly, CPI can be calculated by dividing the total cycles by the total number of instructions. That means we can predict our proxy benchmark's micro-architectural metrics if we know the execution times of each basic block.
\par
In order to make the proxy benchmark's micro-architectural metrics closely match those of the real benchmark and also ensure that the proxy benchmark's running time stays within the specified limit, we need to determine appropriate values for $N_1$,...,$N_m$ that satisfy the constraints of the Equation System \eqref{LM}. Within Equation System \eqref{LM}, the first equation represents multiple equations. In this equation, $metric^{i}$ refers to the ith target micro-architectural metric, which is calculated as the quotient of $f_{A}^{i}$ and $f_{B}^{i}$. If we hope proxy and real benchmarks have more same micro-architectural metrics, we just need to add more equations into Equation System \eqref{LM}, which is very simple. Just by adding equations, our method can deal with the problem that the micro-architectural metric set is huge.
\begin{equation}
\label{LM}
\left\{
\begin{aligned}
\cfrac{\sum_{j=1}^{m} f^{i}_{A}(P_j) \cdot \cfrac{N_j}{N_0}}{\sum_{j=1}^{m} f^{i}_{B}(P_j) \cdot \cfrac{N_j}{N_0}}=metric^{i}, i=1,2,...,q\\
\sum_{j=1}^{m} instructions(P_j) \cdot \cfrac{N_j}{N_0} = ins_1 \\
\end{aligned}
\right.
\end{equation}
The last equation in Equation System \eqref{LM} is used to limit the total number of instructions in the proxy benchmark, and $ins_1$ represents the total number of instructions in the proxy benchmark. Equation System \eqref{LM} can be easily transformed into a linear equation system. Although sometimes the linear equation system doesn't have a unique solution, we can always find the best solution by a non-negative least square method.  
After that, we can construct a proxy benchmark that meets micro-architectural metrics and time limit requirements in theory.
\subsection{Algorithm Flow}
In the process of constructing a proxy benchmark, many basic blocks are redundant. With the number of basic blocks increasing, the mutual influences between them become stronger. Therefore, it's necessary to select a suitable set of basic blocks before formally constructing a proxy benchmark. For a proxy benchmark, we initially consider using all basic blocks to construct and solve the Equation System \eqref{LM}. As a result, some of the basic blocks will have zero execution times, while others will not. It is the latter that we are interested in.
\par
The linear combination method described in \ref{Linear Combination Method} allows us to theoretically construct a proxy benchmark that meets the micro-architectural metrics and time limit requirements. However, in practice, the proxy benchmark may not meet the metric requirements when it actually runs. This can be attributed to several factors, including the mutual influences between the basic blocks, the error of measuring tools, the instability of the system environment, etc. 
\par
We conduct multiple rounds of iteration on the proxy benchmark. During each round, we run the proxy benchmark, measure micro-architectural metrics, and increase the execution times of each basic block based on the measurement results. The algorithm used for this process is provided in Algorithm \ref{alg: Align}. In Equation System \eqref{LM2}, $f_{A}^{i}$ and $f_{B}^{i}$ represent the occurrence times of hardware events of the previous round's proxy benchmark, while $f_{A}^{i}(P_j)$ and $f_{B}^{i}(P_j)$ remain the same meaning as Equation \eqref{LM}. Additionally, we utilize $\Delta N_k$ to denote the increase in execution times of the kth basic block in the present round. $\Delta ins$ represents the total increase of instructions in this round, and the left variables remain the same meaning as Equation System \eqref{LM}. 
Equation System \eqref{LM2} can be easily transformed into a linear system, and we can also find its best solution by the non-negative least square method and ensure the linear equation is astringent.
\begin{algorithm}
    \caption{Align}
    \label{alg: Align}
    \begin{algorithmic}
        \Require The picked basic block set $B$, target metrics, the number of instructions of the first round's proxy benchmark $ins_1$
        \Ensure The proxy benchmark $PB$
        \State $(N_1,N_2,...,N_p) \gets$ solve the Equation System \eqref{LM}
        \State PB $\gets$ construct a proxy benchmark according to $(N_1,N_2,...,N_p)$
        \State run $PB$ and measure its hardware events' occurrence times
        \For{$epoch \gets 2 $ to $10$}
            \State $\Delta ins \gets instructions \times 0.2$ // instructions is the number of instructions of the last round's proxy benchmark
            \State $(\Delta N_1,\Delta N_2,...,\Delta N_p) \gets $ solve the Equation System \eqref{LM2}
            \State $(N_1,N_2,...,N_p) \gets (N_1+\Delta N_1,N_2+\Delta N_2,...,N_p+\Delta N_p)$
            \State PB $\gets$ construct the proxy benchmark according to $(N_1,N_2,...,N_p)$
            \State run $PB$ and measure its hardware events' occurrence times
        \EndFor
        \State \Return $PB$
    \end{algorithmic}
\end{algorithm}
\begin{equation}
\label{LM2}
\left\{
\begin{aligned}
\cfrac{f_{A}^{i}+\sum_{j=1}^{m}{f}_{A}^{i}(P_j) \cdot \cfrac{\Delta N_j}{N_0}}{f_{B}^{i}+\sum_{j=1}^{m} f_{B}^{i}(P_j) \cdot \cfrac{\Delta N_j}{N_0}}=metric^{i}, i=1,2,3...,q\\
\sum_{j=1}^{m} instructions(P_j) \cdot \cfrac{\Delta N_k}{N_0} = \Delta ins \\
\end{aligned}
\right.
\end{equation}
\section{Evaluation}
\label{section: Evaluation}
In this section, we assess the effectiveness of our proxy benchmark generation methodology. 
Many database systems are equipped with stress testing tools, such as db\_bench in RocksDB, Sysbench, and MySQLslap in MySQL. We use these tools to obtain some real benchmarks as listed in Table \ref{Real Benchmarks}.
By employing our methodology, we generate fifteen proxy benchmarks. Subsequently, we measure the accuracy of their micro-architectural metrics as well as their running time.
\begin{table}
\centering
\caption{Real Benchmarks}
\begin{tabular}{|l|l|}
\hline
Benchmark Tool & Workloads \\
\hline
db\_bench & fillrandom, compact, readrandom\\
\hline
Sysbench &  \begin{tabular}[c]{@{}l@{}}oltp\_delete, oltp\_insert, oltp\_point\_select, oltp\_read\_only, \\oltp\_read\_write, oltp\_write\_only, oltp\_update\_non\_index,\\oltp\_update\_index,   select\_random\_points, select\_random\_ranges\end{tabular} \\
\hline
MySQLslap & normal, partition \\
\hline
\end{tabular}
\label{Real Benchmarks}
\end{table}
\subsection{Experiment Setups}
\par
We use a Linux server to conduct our experiments. This server is equipped with two Intel Xeon E5645 processors, each having six physical cores. The operation system is CentOS 6.10 with the Linux kernel version 3.11.40. The server has a total memory of 32GB. The total bandwidth is 72 bits, while the data bandwidth is 64 bits. We use GCC 5.4.0 to compile our proxy benchmarks, and all GCC configurations are set to default.
\subsection{Accuracy}
Equation \eqref{Accuracy Computing} is used to compute the accuracy for each metric in Table \ref{Micro-architectural Metrics}. $Metric_R$ represents the metric value of the real benchmark, and $Metric_P$ represents the metric value of the proxy benchmark. The overall accuracy for a category of metrics is represented by the lowest accuracy among them. 
\begin{equation}
    \label{Accuracy Computing}
    Accuracy(Metric_R,Metric_P)=1-|\frac{Metric_R-Metric_P}{Metric_R}|
\end{equation}
\begin{figure}[htb]
\centering
\includegraphics[scale=0.6]{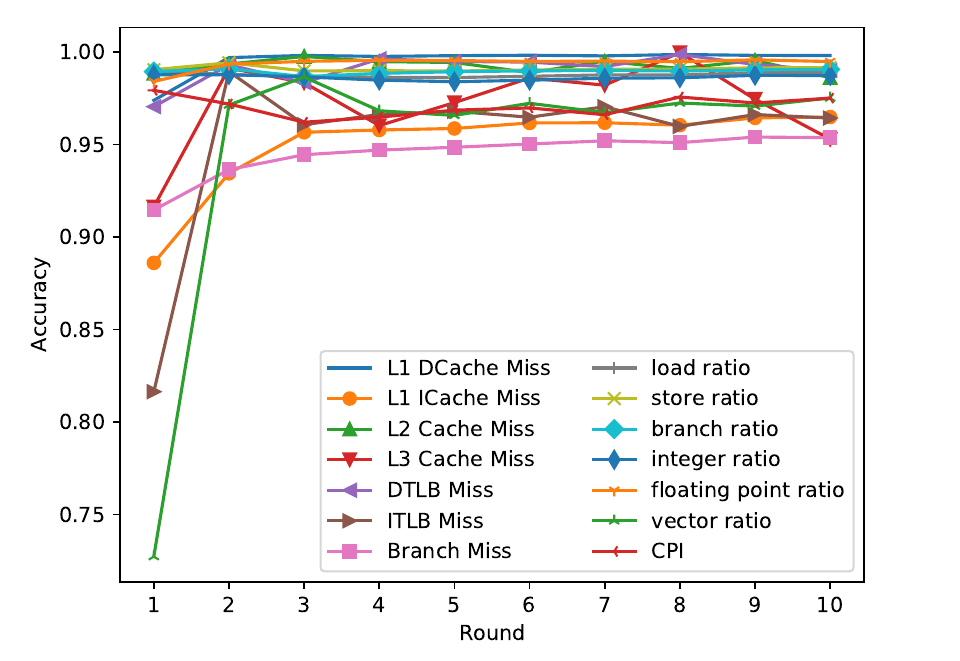}
\caption{A Sample of Constructing Proxy Benchmark.} \label{A Sample of Constructing Proxy Benchmark}
\end{figure}
\par
To construct the proxy benchmark for each real benchmark, we follow the method described in \ref{Linear Combination Method}, which involves a 10-round iteration. In the first round, we create a proxy benchmark containing 0.5 billion instructions. In each subsequent round, the number of instructions is increased by 20\% to correct any metric errors. After 10 rounds of iteration, we obtain the proxy benchmark we require. Fig. \ref{A Sample of Constructing Proxy Benchmark} provides a sample of this process. In the 10th round, the accuracy of every metric is over 95\%.
\begin{figure}[htb]
\centering
\includegraphics[scale=0.3]{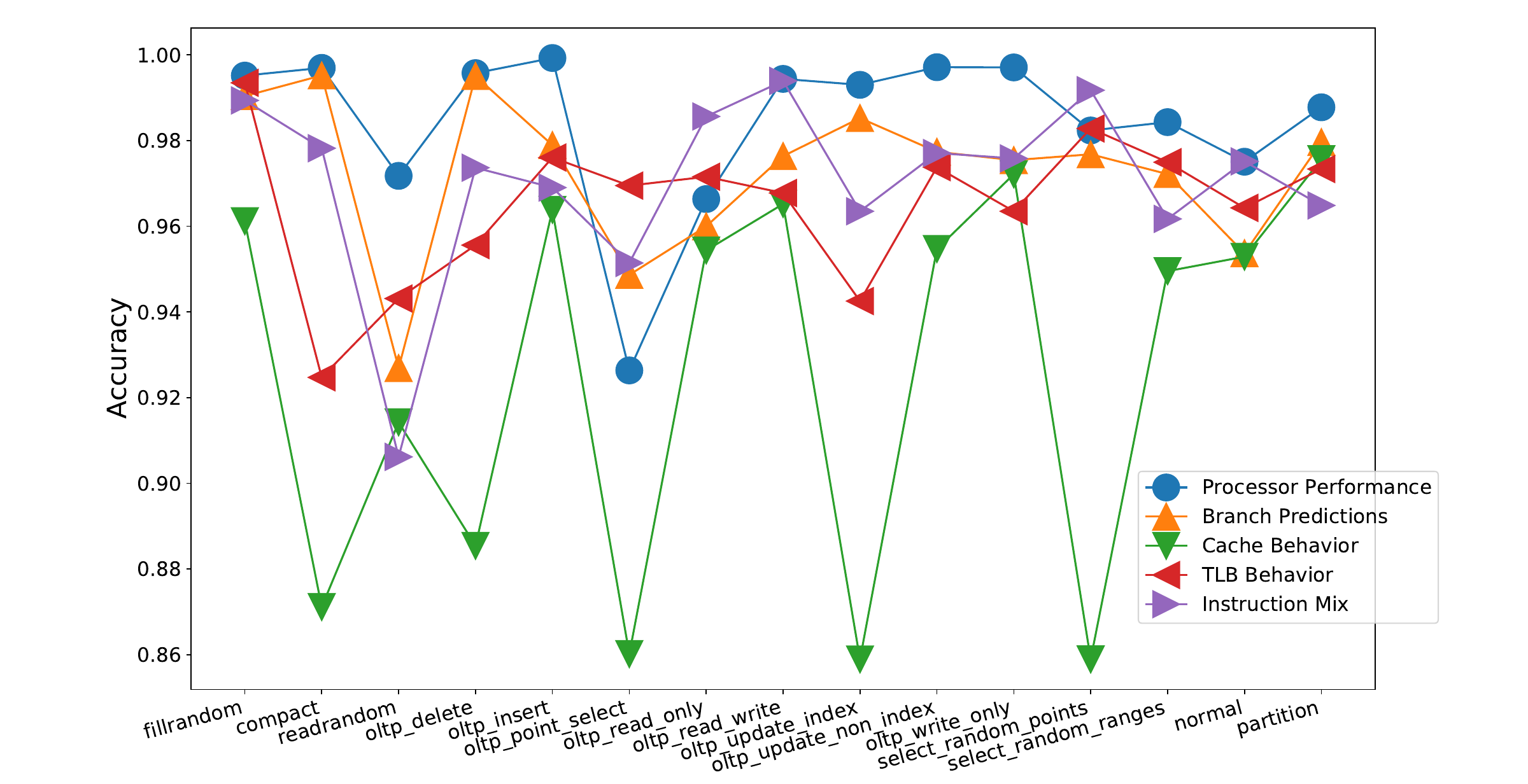}
\caption{Accuracy of All Proxy Benchmarks.} \label{Accuracy of All Proxy Benchmarks}
\end{figure}
\par
\textbf{Processor Performance Accuracy}. Fig. \ref{Accuracy of All Proxy Benchmarks} displays the processor performance accuracy of our proxy benchmarks compared to real benchmarks. The lowest accuracy is 92.6\%, while the average accuracy is 98.4\%. These results prove that our proxy benchmarks closely resemble the corresponding real benchmarks in terms of processor performance.
\par
\textbf{Branch Prediction}. Fig. \ref{Accuracy of All Proxy Benchmarks} shows the branch prediction accuracy of our proxy benchmarks. The minimal accuracy is 92.7\% belonging to 'readrandom' real benchmark. The average accuracy is 97.3\%. 
\par
\textbf{Cache Behavior}. Our proxy benchmarks' cache behavior accuracy can be observed from Fig. \ref{Accuracy of All Proxy Benchmarks}. The lowest accuracy is 85.9\%, and the average accuracy is 92.7\%. It's worth noting that more than half of the proxy benchmarks exhibit an accuracy over 90\%.
\par
\textbf{TLB Behavior}. According to Fig. \ref{Accuracy of All Proxy Benchmarks}, the minimal accuracy observed in TLB behavior is 92.5\%, and the average accuracy is 96.5\%. It is important to note that the majority of the proxy benchmarks demonstrate an accuracy of over 95\%.
\par
\textbf{Instruction Mix}. In terms of instruction mix, according to Fig. \ref{Accuracy of All Proxy Benchmarks}, the minimal accuracy observed is 90.6\%, and the average accuracy is 97.0\%. Most of them have a higher instruction mix accuracy than 95\%. Our proxy benchmarks have similar instruction structures to corresponding real benchmarks.
\par
From the perspectives of metrics,  we can conclude that our proxy benchmarks exhibit similar characteristics to the corresponding real benchmarks. 
\subsection{Running Time}
\begin{figure}
\centering\includegraphics[scale=0.22]{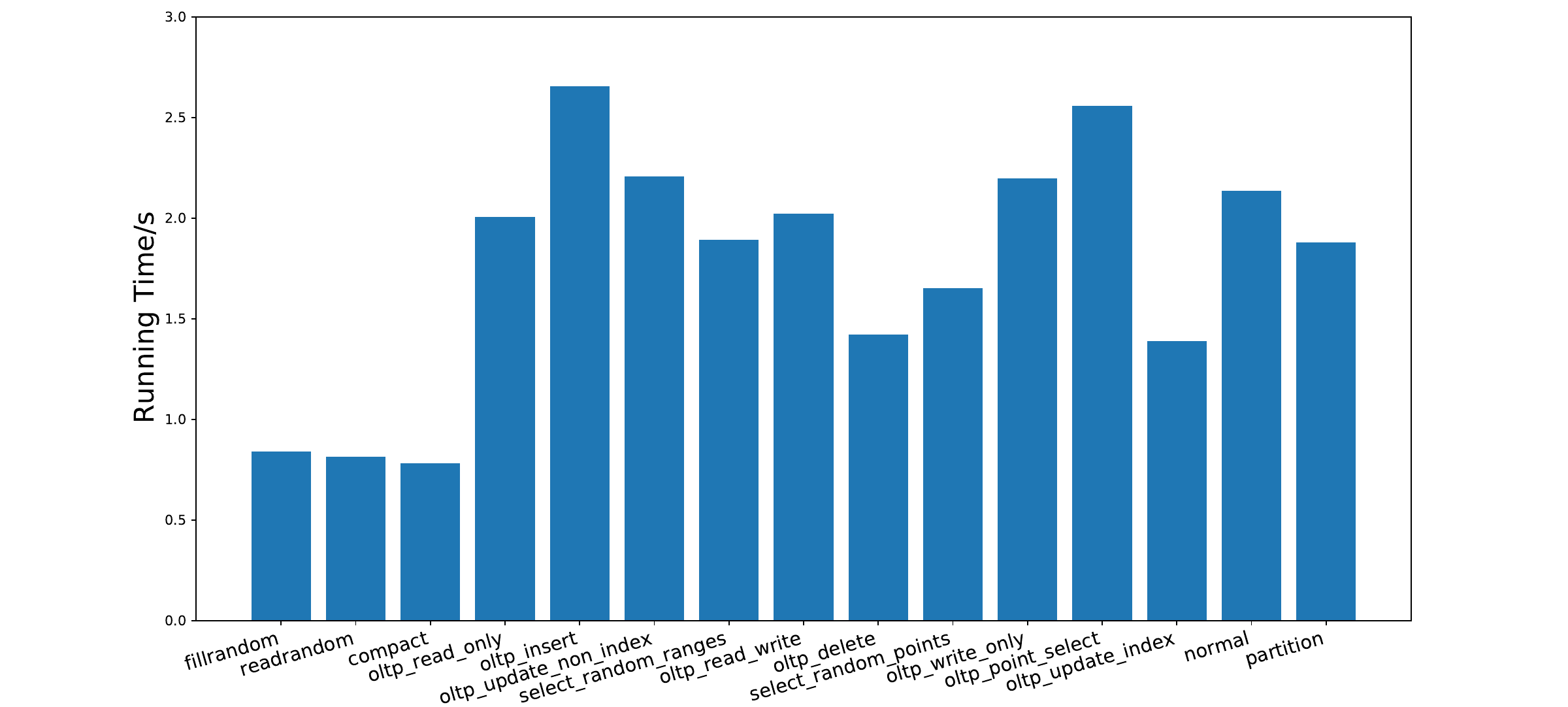}
\caption{Running Time of Proxy benchmarks.} \label{Running Time of Proxy benchmarks}
\end{figure}

We measure the running time of fifteen proxy benchmarks, and the results are displayed in Fig. \ref{Running Time of Proxy benchmarks}. It can be observed that the running time of each db\_bench proxy benchmark is below 1 second, and the running time of each proxy benchmark is below 3 seconds. The average running time of these proxy benchmarks is just 1.62 seconds.
\subsection{Summary}
\par
The average accuracy of every micro-architectural metric between the proxy benchmark and the real benchmark 
is over 92\%, which proves that our proxy benchmarks have almost the same micro-architectural characteristics as real benchmarks. Moreover, the average running time of the proxy benchmarks is just 1.62 seconds. 
Therefore, our proxy benchmark methodology is valid and effective for the micro-architectural metrics.
\section{Case Studies}
\label{section: Case Studies}
\par
In this section, we conduct two use case studies to assess the consistency between real and proxy benchmarks in terms of micro-architectural metrics across different configurations. We will evaluate them from the following perspectives: 1) can proxy and real benchmarks keep consistent when prefetch is enabled? 2) can proxy and real benchmarks keep consistent when Hyper-Threading is enabled? 
\par
To perform these evaluations, we will modify the configurations of our machine. We will rerun both our real benchmarks and proxy benchmarks and measure their respectively micro-architectural metrics. The correlation coefficient will be calculated using Equation \eqref{eq: correlation coefficient}, while the average error will be calculated using Equation \eqref{eq: error}. Here, $x$ represents the micro-architectural metric of the real benchmarks, while $y$ represents the metric of corresponding proxy benchmarks.
\begin{equation}
    \label{eq: correlation coefficient} \rho(x,y)=\cfrac{\sqrt{\sum_{i=1}^{n}(x_i-\overline{x})(y_i-\overline{y})}}{\sqrt{\sum_{i=1}^{n}(x_i-\overline{x})^2 \sum_{i=1}^{n}(y_i-\overline{y})^2}}
\end{equation}
\begin{equation}
    \label{eq: error}
\overline{error(x,y)}=\cfrac{\sum_{i=1}^{n}|\cfrac{y_i-x_i}{x_i}|}{n}
\end{equation}
\subsection{Prefetch Strategy Setting}
\par
Prefetch is a strategy employed by the CPU to fetch data from memory into the cache in advance. The decision to use prefetch can impact the cache behavior and, ultimately, the processor performance. We will focus on cache behavior and processor performance in this part. Fig. \ref{CPI, Branch Miss, DTLB Miss and ITLB Miss When Using Prefetch Strategy} and Fig. \ref{L1 DCache Miss, L1 ICache Miss, L2 Cache Miss and L3 Cache Miss When Using Prefetch Strategy} present the micro-architectural metrics of real and proxy benchmarks when prefetch enabled.
\par
Fig. \ref{CPI, Branch Miss, DTLB Miss and ITLB Miss When Using Prefetch Strategy}a presents the CPI of real benchmarks and corresponding proxy benchmarks. We observe that most of them exhibit similar CPI. Although the largest error can achieve 36\%, the average error remains acceptable, just 7.5\%. Moreover, the real benchmarks and proxy benchmarks display a high correlation($\rho$=0.980). These results illustrate that the proxy benchmarks can retain the processor performance characteristic of the real benchmarks, even when prefetch is enabled.
\begin{figure}[htb]
\centering
\includegraphics[width=0.9\textwidth]{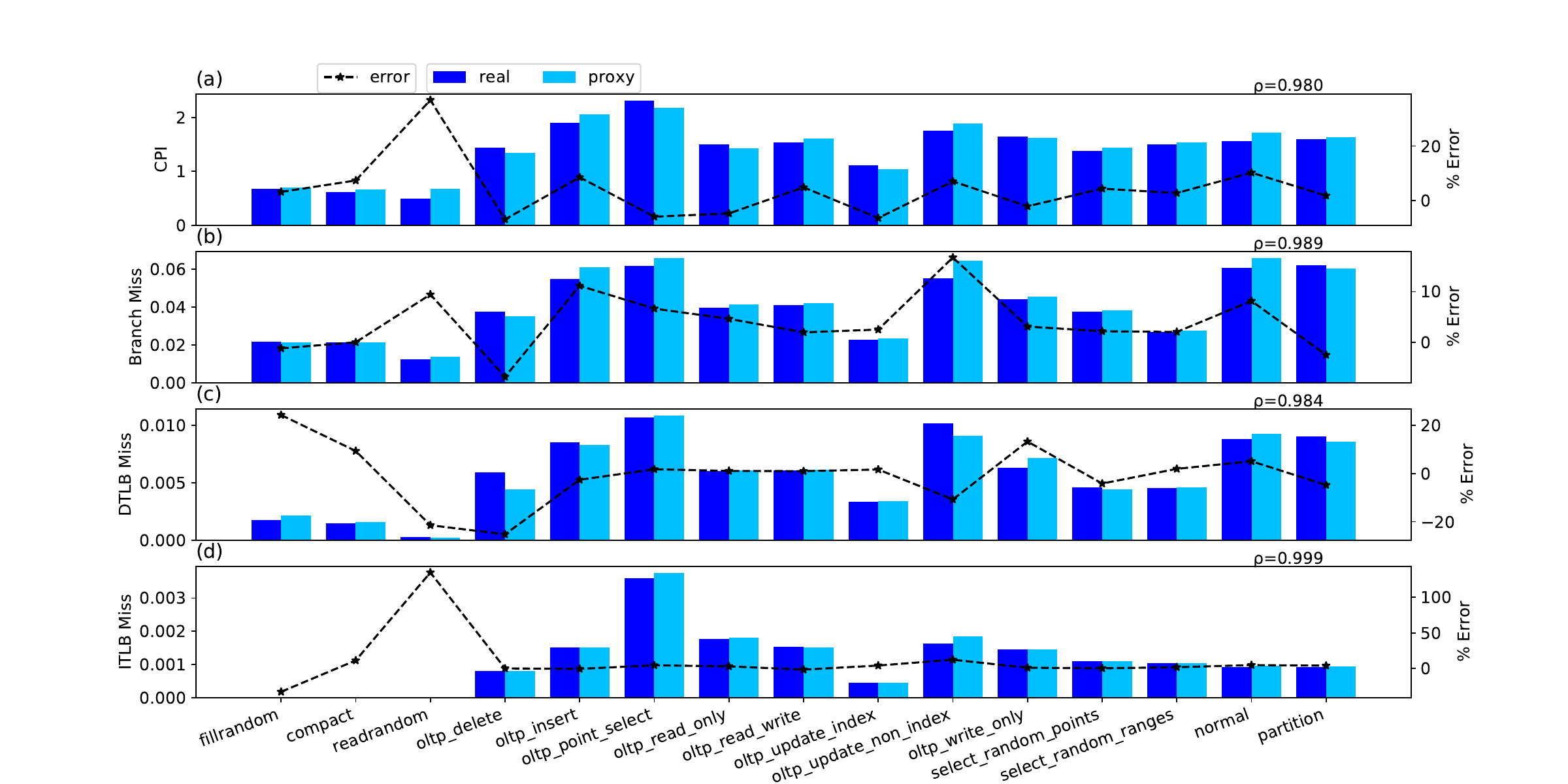}
\caption{CPI, Branch Miss, DTLB Miss, ITLB Miss When Prefetch is enabled.} \label{CPI, Branch Miss, DTLB Miss and ITLB Miss When Using Prefetch Strategy}
\end{figure}
\par
Fig. \ref{L1 DCache Miss, L1 ICache Miss, L2 Cache Miss and L3 Cache Miss When Using Prefetch Strategy}a, Fig. \ref{L1 DCache Miss, L1 ICache Miss, L2 Cache Miss and L3 Cache Miss When Using Prefetch Strategy}b, Fig. \ref{L1 DCache Miss, L1 ICache Miss, L2 Cache Miss and L3 Cache Miss When Using Prefetch Strategy}c and Fig. \ref{L1 DCache Miss, L1 ICache Miss, L2 Cache Miss and L3 Cache Miss When Using Prefetch Strategy}d present the cache behavior of real and proxy benchmarks. The average errors are respectively 4.8\%, 7.0\%, 10.8\%, 43.4\%, and the correlation coefficients are respectively 0.986, 0.993, 0.820, 0.846. We can find that L1 DCache and ICache Miss both have small errors and strong correlations. The average errors of L2 and L3 Cache Miss are bigger, but they still keep correlation at a high level. We can conclude that our proxy benchmarks can retain the cache behavior characteristic of real benchmarks.
For Branch Miss, DTLB Miss, and ITLB Miss, their average errors are respectively 5.26\%, 8.5\%, 14.4\%, and their correlation coefficients are respectively 0.983, 0.984, 0.999. 
\begin{figure}[htb]
\centering
\includegraphics[width=0.9\textwidth]{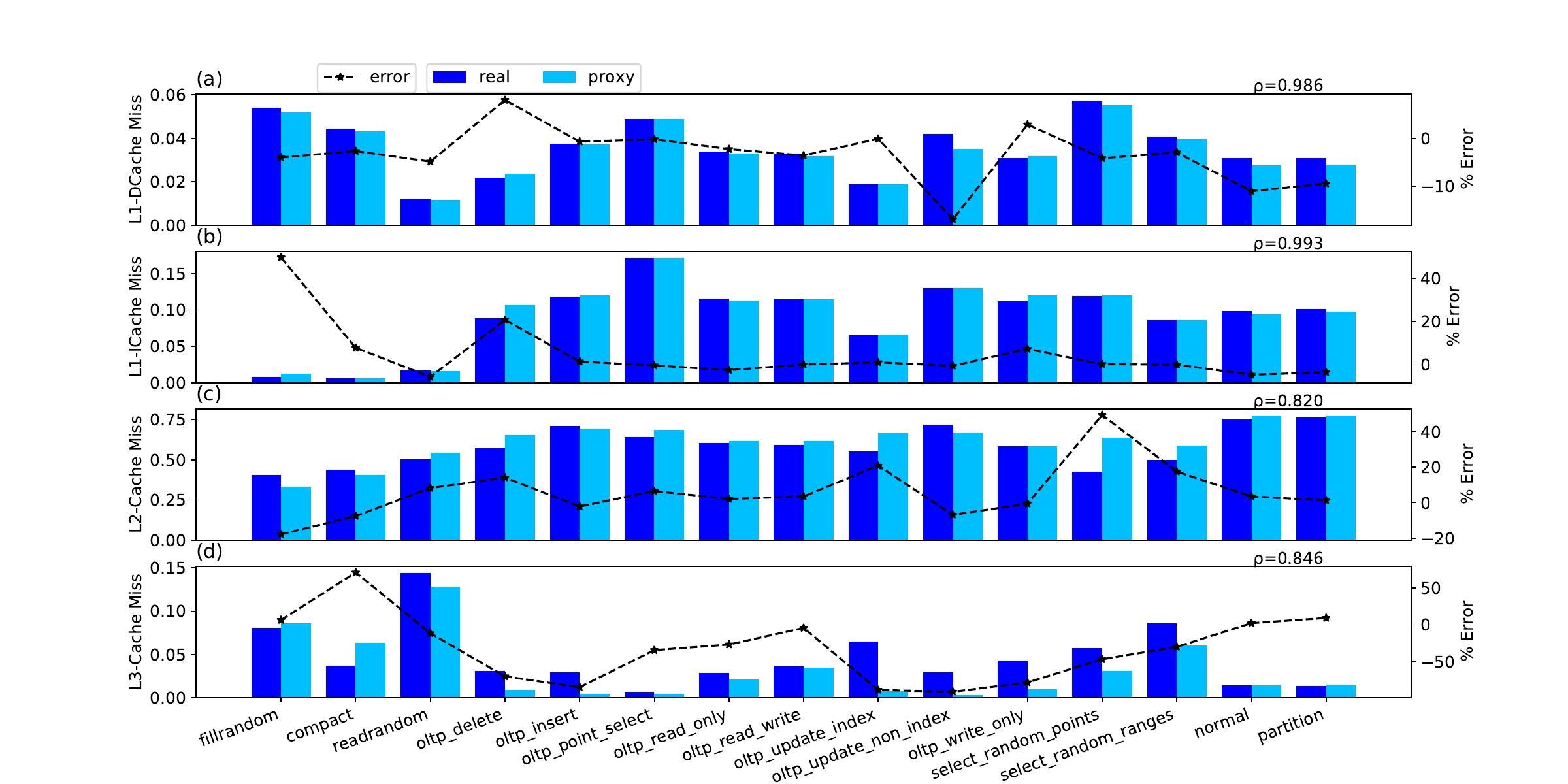}
\caption{Miss of L1 DCache, L1 ICache, L2 Cache, L3 Cache When Prefetch is enabled.} \label{L1 DCache Miss, L1 ICache Miss, L2 Cache Miss and L3 Cache Miss When Using Prefetch Strategy}
\end{figure}
\par 
\subsection{Hyper-Threading Technology}
Hyper-Threading(HT) is Intel's parallel computation technology. HT enhances the parallel computation performance of a CPU by offering two logical threads on one core. Fig. \ref{CPI, Branch Miss, DTLB Miss and ITLB Miss When Using HT Technology} and Fig. \ref{L1 DCache Miss, L1 ICache Miss, L2 Cache Miss and L3 Cache Miss When Using HT Technology} present the metrics of real benchmarks and proxy benchmarks when running with HT enabled on the server.
\par
Fig. \ref{CPI, Branch Miss, DTLB Miss and ITLB Miss When Using HT Technology}a compares the CPI between real benchmarks and proxy benchmarks. A strong correlation($\rho$=0.876) is observed between the CPI 
of real and proxy benchmarks. However, there are significant gaps in some cases, such as select\_random\_ranges, which has an error of 31.4\%. It can be observed that the CPI is quite close for the first three real benchmarks and proxy benchmarks but not for the others. This can be explained by the fact that only the first three real benchmarks are single-thread programs, while all proxy benchmarks are single-thread programs. Therefore, our proxy benchmarks can retain single-thread real benchmarks' properties better. In addition to CPI, as shown in Fig. \ref{CPI, Branch Miss, DTLB Miss and ITLB Miss When Using HT Technology}b, similar patterns can be observed in the results of Branch Miss. Branch Miss between the first three real benchmarks and proxy benchmarks are quite close. This may be due to the impact of frequent thread switches on branch prediction.
\begin{figure}[htb]
\centering
\includegraphics[width=0.9\textwidth]{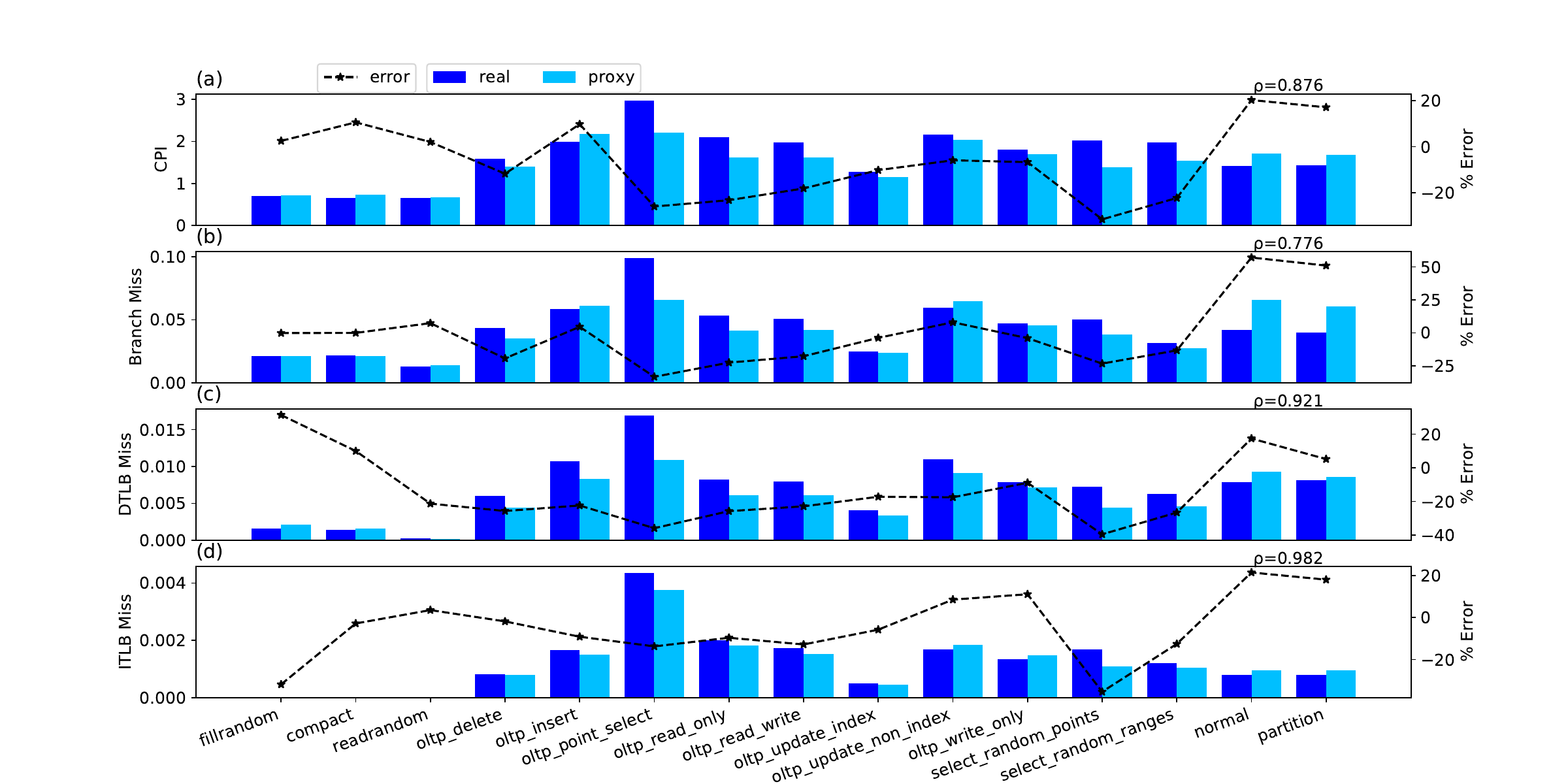}
\caption{CPI, Branch Miss, DTLB Miss, ITLB Miss When HT is enabled.} \label{CPI, Branch Miss, DTLB Miss and ITLB Miss When Using HT Technology}
\end{figure}
\par
Fig. \ref{L1 DCache Miss, L1 ICache Miss, L2 Cache Miss and L3 Cache Miss When Using HT Technology}a, Fig. \ref{L1 DCache Miss, L1 ICache Miss, L2 Cache Miss and L3 Cache Miss When Using HT Technology}b, Fig. \ref{CPI, Branch Miss, DTLB Miss and ITLB Miss When Using HT Technology}a and Fig. \ref{CPI, Branch Miss, DTLB Miss and ITLB Miss When Using HT Technology}b prove proxy benchmarks retain real benchmarks' properties well in L1 DCache, L1 ICache, DTLB and ITLB. The correlation coefficients are respectively 0.950, 0.982, 0.921, and 0.982. The average errors are respectively 10.2\%, 7.4\%, 21.8\%, and 13.2\%. The results of L2 Cache and L3 Cache are worse, while the average errors are 16.2\% and 25.5\%, and the correlation coefficients are 0.800 and 0.991. 
\begin{figure}
\centering
\includegraphics[width=0.9\textwidth]{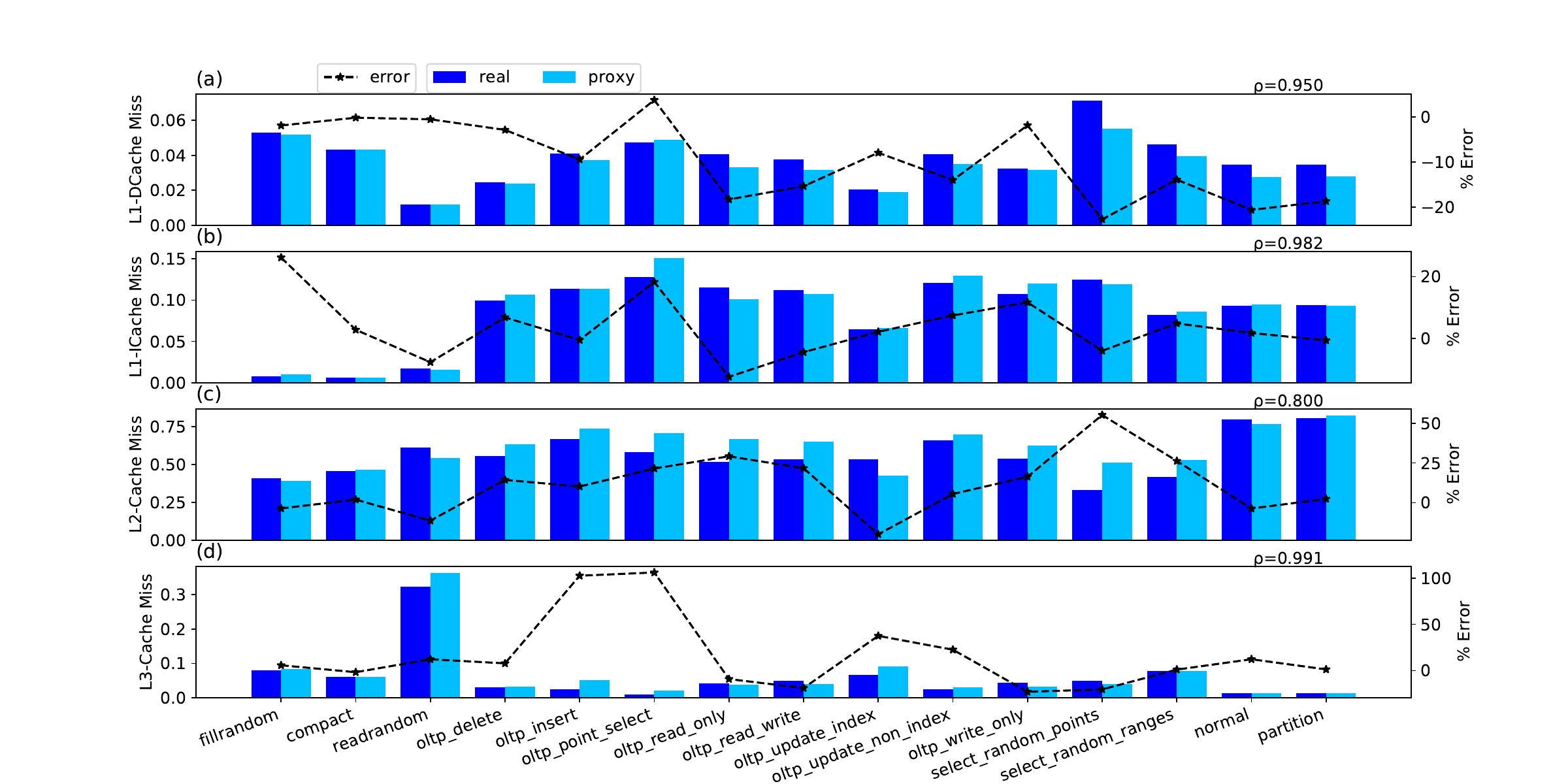}
\caption{Miss of L1 DCache, L1 ICache, L2 Cache, L3 Cache When HT is enabled.} \label{L1 DCache Miss, L1 ICache Miss, L2 Cache Miss and L3 Cache Miss When Using HT Technology}
\end{figure}
\par
When HT is enabled, the correlation coefficients of most micro-architectural metrics keep at a high level, which demonstrates that our proxy benchmarks can keep consistent with real benchmarks. For CPI and Branch Miss, the decrease in correlation can be attributed to our methodology's shortcoming that the proxy benchmark can't retain the real benchmark's multi-thread characteristic well.
\subsection{Summary}
\par
When prefetch or HT is enabled, most micro-architectural metrics' correlation coefficients are high between real and proxy benchmarks. Therefore, the micro-architectural metrics of real and proxy benchmarks can keep consistent before and after prefetch or HT is enabled. This implies that our proxy benchmark methodology could be used for micro-architectural design evaluations.
\section{Related Work}
\label{section: Related Work}
Many big data benchmarks have been proposed to evaluate big data system performance. Such as BigBench\cite{BigBench}, TPC-DS\cite{TPC-DS}, BigDataBench\cite{gao2018bigdatabench,BigDataBenchWebSearchEngines,BigDataBenchInternetServices}, CloudSuite\cite{CloudSuite}, and YCSB\cite{YCSB}.
However, it's difficult for these benchmarks to run on simulators because of their complex software stacks and long running time.
\par
Reducing data size is an effective way to reduce running time for workloads. Keeton et al. successfully reduce workload running time by replacing complex queries with simple ones, thus reducing data size\cite{keeton2000towards}. Shao et al. achieve similar goals by modifying queries, reducing the number of concurrent clients, and overall data size for DSS and OLTP workloads\cite{Shao}. Barroso et al. and Ertvelde et al. opt for using partial datasets instead of the entire dataset\cite{BarrosoGB98,ErtveldeE10}, which also proved to be a viable method for reducing running time. Although these methods effectively reduce running time, they still depend on complex software stacks.
\par
Extracting workload segments is another popular way. \cite{DBLP:conf/iccd/ConteHM96,DBLP:conf/isca/WunderlichWFH03,DBLP:conf/date/LuJTL11} sample the instruction stream and fuse the sampled segments into new programs. Kernel benchmarks, consisting  of a set of kernels
extracted from real application~\cite{QuantitativeApproach}, such as \cite{NPB}, are widely used in high-performance computing. SimPoint~\cite{SimPoint} method uses BBV~\cite{BBV} to select some basic blocks to represent the overall program. However, these methods also need the support of complex software stacks.
\par
The last way is to construct proxy benchmarks~\cite{Cloudmix,PerfProx,data-motif,data-motif-proxy}. This method involves using small program segments to construct proxy benchmarks, which can then be used to replace the real benchmarks while can both reduce running time and get rid of complex software stacks.
\section{Conclusion}
\label{section: Conclusion}
In this paper, based on simple program fragments called basic blocks, we propose a novel proxy benchmark generation methodology. 
This methodology involves linearly combining basic blocks and adjusting their execution times through iteration to mimic real benchmarks. 
Our case studies also demonstrate that 
our proxy benchmarks are consistent with real benchmarks before and after prefetch or Hyper-Threading is turned on.


\begin{thebibliography}{10}
\providecommand{\url}[1]{\texttt{#1}}
\providecommand{\urlprefix}{URL }
\providecommand{\doi}[1]{https://doi.org/#1}

\bibitem{DatabaseResearch}
Indrawan{-}Santiago, M.: Database research: Are we at a crossroad? Reflection
  on NoSQL. In: 2012 15th International Conference on Network-Based Information
  Systems. pp. 45--51. {IEEE} Computer Society (2012)

\bibitem{MongodbvsOracle}
Boicea, A., Radulescu, F., Agapin, L.I.: MongoDB vs Oracle - database
  comparison. In: 2012 Third International Conference on Emerging Intelligent
  Data and Web Technologies. pp.
  330--335. {IEEE} Computer Society (2012)

\bibitem{LargeScaleDataAnalysis}
Pavlo, A., Paulson, E., Rasin, A., Abadi, D.J., DeWitt, D.J., Madden, S.,
  Stonebraker, M.: A comparison of approaches to large-scale data analysis. In:
  Proceedings of the 2009 {ACM} {SIGMOD} International Conference on Management of
  Data. pp. 165--178. {ACM} (2009)

\bibitem{SQLvsNoSQL}
Li, Y., Manoharan, S.: A performance comparison of SQL and NoSQL databases. In:
  2013 IEEE Pacific Rim Conference on Communications, Computers and Signal Processing (PACRIM). pp. 15--19. IEEE (2013)

\bibitem{SPECCPU2006}
SPECCPU2017: https://www.spec.org/cpu2017/

\bibitem{PARSEC}
Bienia, C.: Benchmarking modern multiprocessors. Princeton University (2011)

\bibitem{CloudSuite}
Ferdman, M., Adileh, A., Ko{\c{c}}berber, Y.O., Volos, S., Alisafaee, M.,
  Jevdjic, D., Kaynak, C., Popescu, A.D., Ailamaki, A., Falsafi, B.: Clearing
  the clouds: A study of emerging scale-out workloads on modern hardware. In:
  Proceedings of the 17th International Conference on Architectural Support for
  Programming Languages and Operating Systems (ASPLOS). pp. 37--48. {ACM} (2012)

\bibitem{gao2018bigdatabench}
Gao, W., Zhan, J., Wang, L., Luo, C., Zheng, D., Wen, X., Ren, R., Zheng, C.,
  He, X., Ye, H., et~al.: BigDataBench: A scalable and unified big data and AI
  benchmark suite. arXiv preprint arXiv:1802.08254 (2018)

\bibitem{BigDataBenchWebSearchEngines}
Gao, W., Zhu, Y., Jia, Z., Luo, C., Wang, L., Li, Z., Zhan, J., Qi, Y., He, Y.,
  Gong, S., Li, X., Zhang, S., Qiu, B.: BigDataBench: A big data benchmark
  suite from web search engines. arXiv preprint arXiv:1307.0320 (2013)

\bibitem{BigDataBenchInternetServices}
Wang, L., Zhan, J., Luo, C., Zhu, Y., Yang, Q., He, Y., Gao, W., Jia, Z., Shi,
  Y., Zhang, S., Zheng, C., Lu, G., Zhan, K., Li, X., Qiu, B.: BigDataBench:
  {A} big data benchmark suite from internet services. In: 2014 {IEEE} 20th International Symposium on High Performance Computer Architecture (HPCA). pp. 488--499. {IEEE} Computer Society (2014)

\bibitem{Cloudmix}
Han, R., Zong, Z., Zhang, F., V{\'{a}}zquez{-}Poletti, J.L., Jia, Z., Wang, L.:
  Cloudmix: Generating diverse and reducible workloads for cloud systems. In:
  2017 {IEEE} 10th International Conference on Cloud Computing (CLOUD). pp.
  496--503. {IEEE} Computer Society (2017)

\bibitem{PerfProx}
Panda, R., John, L.K.: Proxy benchmarks for emerging big-data workloads. In:
  Proceedings of the 26th International Conference on Parallel Architectures and Compilation Techniques (PACT). pp. 105--116. {IEEE} Computer Society (2017)

\bibitem{data-motif}
Gao, W., Zhan, J., Wang, L., Luo, C., Zheng, D., Tang, F., Xie, B., Zheng, C.,
  Wen, X., He, X., Ye, H., Ren, R.: Data motifs: A lens towards fully
  understanding big data and {AI} workloads. In: Proceedings of the 27th International Conference on Parallel Architectures and Compilation Techniques (PACT). {ACM} (2018) 

\bibitem{data-motif-proxy}
Gao, W., Zhan, J., Wang, L., Luo, C., Jia, Z., Zheng, D., Zheng, C., He, X.,
  Ye, H., Wang, H., Ren, R.: Data motif-based proxy benchmarks for big data and
  {AI} workloads. In: Proceedings of the 2018 IEEE International Symposium on Workload Characterization (IISWC). pp. 48--58. {IEEE} Computer Society (2018)

\bibitem{IntelDeveloperManul}
Intel: https://perfmon-events.intel.com/

\bibitem{Perf}
Perf: https://perf.wiki.kernel.org/index.php/Main Page

\bibitem{BigBench}
Ghazal, A., Rabl, T., Hu, M., Raab, F., Poess, M., Crolotte, A., Jacobsen, H.:
  BigBench: Towards an industry standard benchmark for big data analytics. In: Proceedings of the 2013 {ACM}
  {SIGMOD} International Conference on Management of Data. pp. 1197--1208. {ACM} (2013)

\bibitem{TPC-DS}
P{\"{o}}ss, M., Smith, B., Koll{\'{a}}r, L., Larson, P.: TPC-DS, taking
  decision support benchmarking to the next level. In: Proceedings of the 2002 {ACM} {SIGMOD} International
  Conference on Management of Data.
  pp. 582--587. {ACM} (2002)

\bibitem{YCSB}
Dey, A., Fekete, A.D., Nambiar, R., R{\"{o}}hm, U.: {YCSB+T:} Benchmarking
  web-scale transactional databases. In: 2014 IEEE 30th International Conference on Data Engineering Workshops (ICDE). pp. 223--230. {IEEE} Computer Society
  (2014)

\bibitem{keeton2000towards}
Keeton, K., Patterson, D.A.: Towards a simplified database workload for
  computer architecture evaluations. Workload Characterization for Computer
  System Design pp. 49--71 (2000)

\bibitem{Shao}
Minglong, Ailamaki, A., Falsafi, B.: DBmbench: Fast and accurate database
  workload representation on modern microarchitecture. In: Proceedings of the
  2005 Conference of the Centre for Advanced Studies on Collaborative Research.
  pp. 254--267. {IBM} (2005)

\bibitem{BarrosoGB98}
Barroso, L.A., Gharachorloo, K., Bugnion, E.: Memory system characterization of
  commercial workloads. In: Proceedings of the 25th Annual International
  Symposium on Computer Architecture. pp. 3--14. {IEEE} Computer Society (1998)

\bibitem{ErtveldeE10}
Ertvelde, L.V., Eeckhout, L.: Benchmark synthesis for architecture and compiler
  exploration. In: Proceedings of the 2010 {IEEE} International Symposium on
  Workload Characterization (IISWC). pp. 1--11. {IEEE} Computer Society (2010)

\bibitem{DBLP:conf/iccd/ConteHM96}
Conte, T.M., Hirsch, M.A., Menezes, K.N.: Reducing state loss for effective
  trace sampling of superscalar processors. In: 1996 International Conference
  on Computer Design {(ICCD} '96), {VLSI} in Computers and Processors. pp.
  468--477. {IEEE} Computer Society (1996)

\bibitem{DBLP:conf/isca/WunderlichWFH03}
Wunderlich, R.E., Wenisch, T.F., Falsafi, B., Hoe, J.C.: {SMARTS:} Accelerating
  microarchitecture simulation via rigorous statistical sampling. In: 30th
  International Symposium on Computer Architecture (ISCA). pp. 84--95.
  {IEEE} Computer Society (2003)

\bibitem{DBLP:conf/date/LuJTL11}
Lu, F., Joseph, R., Trajcevski, G., Liu, S.: Efficient parameter variation
  sampling for architecture simulations. In: Design, Automation and Test in
  Europe (DATE). pp. 1578--1583. {IEEE} (2011)

\bibitem{QuantitativeApproach}
Hennessy, J.L., Patterson, D.A.: Computer Architecture - {A} Quantitative
  Approach, 5th Edition. Morgan Kaufmann (2012)

\bibitem{NPB}
Bailey, D.H., Barszcz, E., Barton, J.T., Browning, D.S., Carter, R.L., Dagum,
  L., Fatoohi, R.A., Frederickson, P.O., Lasinski, T.A., Schreiber, R., Simon,
  H.D., Venkatakrishnan, V., Weeratunga, S.: The NAS parallel benchmarks. Int.
  J. High Perform. Comput. Appl.  \textbf{5}(3),  63--73 (1991)

\bibitem{SimPoint}
Sherwood, T., Perelman, E., Hamerly, G., Calder, B.: Automatically
  characterizing large scale program behavior. In: Proceedings of the 10th
  International Conference on Architectural Support for Programming Languages
  and Operating Systems (ASPLOS-X). pp. 45--57. {ACM} Press (2002)

\bibitem{BBV}
Sherwood, T., Perelman, E., Calder, B.: Basic block distribution analysis to
  find periodic behavior and simulation points in applications. In: Proceedings of the 10th International Conference on Parallel Architectures and Compilation Techniques (PACT). pp. 3--14. {IEEE}
  Computer Society (2001)

\end{thebibliography}
\end{document}